\documentclass[12pt]{iopart}

\usepackage{iopams}
\expandafter\let\csname equation*\endcsname\relax
\expandafter\let\csname endequation*\endcsname\relax
\usepackage{amsmath}
\usepackage{hyperref}
\usepackage{orcidlink}
\usepackage{tabularx}
\usepackage{soul}
\usepackage{multirow}
\usepackage{xcolor}
\usepackage{adjustbox}
\usepackage{cite}
\usepackage[bottom]{footmisc}
\usepackage{changepage}

\begin{document}

\title[ParamANN]{ParamANN: A Neural Network to Estimate Cosmological Parameters for $\Lambda$CDM Universe Using Hubble Measurements}

\author{Srikanta Pal$^{1,a}$\orcidlink{0000-0002-9502-8510} and Rajib Saha$^{1,b}$\orcidlink{0000-0002-4444-1081}\footnote{Corresponding author}}

\address{$^1$Department of Physics, Indian Institute of Science Education and Research Bhopal, Bhopal-462066, Madhya Pradesh, India}
\eads{\mailto{$^a$srikanta18@iiserb.ac.in}, \mailto{psrikanta357@gmail.com}}
\ead{$^b$rajib@iiserb.ac.in}
\vspace{10pt}
\begin{indented}
\item[]October 2024
\end{indented}

\begin{abstract}
In this article, we employ a machine learning (ML) approach for the estimations of four fundamental parameters, namely, the Hubble constant ($H_0$), matter ($\Omega_{0m}$), curvature ($\Omega_{0k}$) and vacuum ($\Omega_{0\Lambda}$) densities of non-flat $\Lambda$CDM model. We use $31$ Hubble parameter values measured by \textit{differential ages} (DA) technique in the redshift interval $0.07 \leq z \leq 1.965$. We create an \textit{artificial neural network} (ParamANN) and train it with simulated values of $H(z)$ using various sets of $H_0$, $\Omega_{0m}$, $\Omega_{0k}$, $\Omega_{0\Lambda}$ parameters chosen from different and sufficiently wide prior intervals. We use a correlated noise model in the analysis. We demonstrate accurate validation and prediction using ParamANN. ParamANN provides an excellent cross-check for the validity of the $\Lambda$CDM model. We obtain $H_0 = 68.14 \pm 3.96$ $\rm{kmMpc^{-1}s^{-1}}$, $\Omega_{0m} = 0.3029 \pm 0.1118$, $\Omega_{0k} = 0.0708 \pm 0.2527$ and $\Omega_{0\Lambda} = 0.6258 \pm 0.1689$ by using the trained network. These parameter values agree very well with the results of global CMB observations of the Planck collaboration. We compare the cosmological parameter values predicted by ParamANN with those obtained by the MCMC method. Both the results agree well with each other. This demonstrates that ParamANN is an alternative and complementary approach to the well-known Metropolis-Hastings algorithm for estimating the cosmological parameters by using Hubble measurements.
\end{abstract}
%
\vspace{2pc}

\section{Introduction}
\label{sec:intro}
Measurements of Hubble expansion rate, $H(z)$, across various redshifts ($z$), offer a valuable tool to study the Universe's evolution due to the relatively simple relationship between the $H(z)$ and fundamental parameters of cosmology. This has the potential to lead to easier interpretation of the results of an analysis by unambiguously identifying the underlying cosmological model. Although a simple 6 parameter $\Lambda$ cold dark matter ($\Lambda$CDM) model is generally believed to outweigh other cosmological models, some observations suggest potential discrepancies between the $\Lambda$CDM model and certain experimental data. The area of incompatibility is known as the so-called problem of `Hubble tension'. The tension was reported since $H_0$ values estimated by local and global observations do not agree with each other. The value of $H_0$ constrained by Planck collaboration~\cite{Planck_2018} shows $\sim3.6\sigma$ tension with the same measured by local observational data, in which local measurements of $H_0$ demand the higher value of this parameter. Recent local observations of Hubble Space Telescope(HST)~\cite{Riess_2018,Riess_2019,Riess_2020,Riess_2021,Riess_2022} estimate the value of $H_0$ which shows approximately $4$-$5\sigma$ tension with Planck's estimation of $H_0$. Local measurement of $H_0$ by using gravitationally lensed quasars, i.e. H0LiCOW~\cite{Wong_2019}, shows $\sim5.3\sigma$ tension with Planck's $H_0$ value. Di Valentino (2021)~\cite{Valentino_2021} estimated the value of $H_0$ by combining 23 local measurements of this parameter and this estimation shows $5.9\sigma$ tension with $H_0$ constrained by CMB observations~\cite{Planck_2018}. \textit{An alternative method to estimate $H_0$ value (along with other cosmological parameters) is of utmost importance in contemporary cosmological analysis.}

It is worth mentioning that the problem of Hubble tension seems to be somewhat perplexing in nature. On one hand, some observations~\cite{Jaeger_2022} indicate a high value of $H_0$. Riess \etal (2022)~\cite{Riess_2022} constrain $H_0 = 73.04$ $\rm{kmMpc^{-1}s^{-1}}$ with a uncertainty of 1.04 $\rm{kmMpc^{-1}s^{-1}}$ using Cepheid-SNe samples for their baseline redshift range $0.0233 < z < 0.15$. The Pantheon+ analysis~\cite{Brout_2022} use 1550 distinct type Ia Supernovae in the redshift interval 0.001 to 2.26 and report the $H_0 = 73.5 \pm 1.1$ $\rm{kmMpc^{-1}s^{-1}}$. Thus it appears that the Hubble tension extends up to a redshift of 2.26 if Supernovae data is used. Interestingly, using the red giant branch (TRGB) calibration to a sample of Type Ia Supernovae, Freedman (2021)~\cite{Freedman_2021} reports $H_0 = 69.8 \pm 0.6$ (stat) $\pm 1.6$ (sys) $\rm{kmMpc^{-1}s^{-1}}$ which is consistent with the $H_0$ value estimated by Planck collaboration~\cite{Planck_2018}. Kelly \etal (2023)~\cite{Kelly_2023} obtain $H_0 = 64.8^{+4.4}_{-4.3}$ $\rm{kmMpc^{-1}s^{-1}}$ by using eight lens models and $66.6^{+4.1}_{-3.3}$ $\rm{kmMpc^{-1}s^{-1}}$ from two preferred models. A recent article~\cite{Mukherjee_2020} report $H_0 = 67.6^{+4.3}_{-4.2}$ $\rm{kmMpc^{-1}s^{-1}}$ using VLBI and gravitational wave observations from bright binary black hole GW190521 and conclude that the value is consistent with the Planck observations, relaxing the tension. At the backdrop of such diverse indications from the observations, it is a very important task to measure the cosmological parameters in an as much general framework (i.e., measuring all the independent cosmological parameters) as possible using the favored $\Lambda$CDM model and new observational data. Leaf \& Melia (2017)~\cite{Leaf_2017} analyze only \textit{differential ages} (DA) Hubble measurements using a two-point diagnostic for model comparison since the measurements of $H(z)$ from cosmic chronometers are model independent. Geng \etal (2018)~\cite{Geng_2018} use $H(z)$ data measured by DA and \textit{baryon acoustic oscillation} (BAO) techniques to constrain the cosmological parameters as well as quantify the impact of future $H(z)$ measurements in the estimation of these parameters. The recent literatures~\cite{Yu_2018,Gomez_2018,Ryan_2018,Ryan_2019,
Cao_2021,Cao_2022} effectively use the low redshift data (i.e., DA+BAO Hubble data, QSO angular size, Pantheon, DES supernova etc.) to analyze the various cosmological models. Almeida \etal (2017)~\cite{Almeida_2017} investigated the inflationary phase and other stages of the Universe within a redshift-dependent Lorentz-violating time-like background and noticed that the dark energy era at low redshifts is dominated by this background. Aktaş \etal (2012)~\cite{Aktas_2012} demonstrated that within the framework of $f(R)$ gravity, dark energy, modeled as both a perfect fluid and a mesonic scalar field in Marder space-time, exhibits phantom-like behavior with an equation of state $\omega<-1$. Iqbal \& Fawad (2019)~\cite{Iqbal_2019} examined Tsallis, Renyi, and Sharma–Mittal holographic dark energy models in a flat Friedmann-Robertson-Walker (FRW) universe, incorporating interactions between dark energy and cold dark matter under the DGP braneworld framework. They have shown that these models agree with the accelerated expansion and stability, consistent with observational data and the $\Lambda$CDM model. In this article, we use $H(z)$ observations alone to constrain the parameters of $\Lambda$CDM model using artificial intelligence (AI) as the driving engine.

We use AI to estimate the Hubble constant ($H_0$) and today's density parameters ($\Omega_{0m}$, $\Omega_{0k}$ and $\Omega_{0\Lambda}$) for $\Lambda$CDM Universe from the $H(z)$ values measured by DA technique~\cite{Jimenez_2002}. We create an ANN (hereafter ParamANN) for modeling a direct mapping function between the observed $H(z)$ and the corresponding four parameters of the $\Lambda$CDM Universe. The density parameters estimated by ParamANN indicate the spatially flat $\Lambda$CDM Universe which is consistent with the results of  Planck collaboration~\cite{Planck_2018}. Moreover, the Hubble constant predicted by ParamANN agrees excellently with Planck's estimation of the same. Furthermore, we compare the parameter values predicted by ParamANN with the same estimated by the Markov chain Monte Carlo (MCMC) approach, e.g. Metropolis-Hastings algorithm~\cite{Hastings_1970}. We notice that both approaches provide similar values with similar levels of uncertainty for each parameter. Therefore, ParamANN can be used as an alternative to the traditional approach for estimating the fundamental parameters from Hubble data. We note that whether the MCMC analysis depends on the specific likelihood function, our efficient and simple neural network (for estimating cosmological parameters) performs as a likelihood-free inference method~\cite{Alsing_2018}.

The primary motivations of our current article stem from the perspective of both theoretical and observational fronts. From the observational side, the problem like Hubble tension~\cite{Planck_2018,Riess_2018,Riess_2019,Riess_2020,Riess_2021,
Riess_2022,Wong_2019,Valentino_2021,Jaeger_2022,Brout_2022} exists and needs further understanding using various types of available data. AI has become one of the promising tools for investigating the observed data since an ML model can predict (in principle) a complicated function once it has been trained successfully. Moreover, there are several articles in the literature~\cite{Macaulay_2013,
Raveri_2016,Shafieloo_2012,Sahni_2014,Zheng_2016,Linder_2003,Shahalam_2015,Leaf_2017,Geng_2018,Gomez_2018,Bengaly_2023,Liu_2019,Arjona_2020,Mukherjee_2022,Garcia_2023} which try to measure the cosmological parameters assuming flat spatial curvature of our Universe. However, it is also important to ask whether we can estimate these cosmological parameters by relaxing the assumption of the flatness of the Universe. In this article, we motivate ourselves to estimate cosmological density parameters along with today's Hubble parameter value for a general non-flat $\Lambda$CDM Universe. From the observational perspective, many new experiments are being proposed, e.g., Echo (aka CMB-Bharat\footnote{\url{http://cmb-bharat.in/}}), CCAT-prime~\cite{Stacey_2018}, PICO~\cite{Hanany_2019}, Lite-Bird~\cite{Hazumi_2020}, SKA~\cite{Dewdney_2009}, which has lower noise level implying the model parameters of a theory can be measured with higher accuracy. The higher accuracy of the future generation observations demands accurate constraining of cosmological parameters to distinguish between different models of the Universe which is a major step to finding a better and more accurate theoretical understanding of the physics of the Universe.

In the modern era, ML techniques are utilized as powerful tools to analyze the observational data of several fields in cosmology~\cite{Ntampaka_2021,Olvera_2022}. Instead of the Metropolis-Hastings algorithm, an artificial neural network (ANN) can be used as an alternative approach for the Bayesian inference in cosmology effectively reducing the computational time~\cite{Graff_2012,Moss_2020,
Hortua_2020,Gomez_2021}. Moreover, ANNs can be performed for the model-free reconstructions of the functions related to cosmology~\cite{Escamilla_2020,Wang_2020a,Dialektopoulos_2022,Gomez_2023}. For the parameter estimation of cosmology by using CMB data, Mancini \etal (2022)~\cite{Mancini_2022} implement several types of ANN and show less computational time in the Bayesian process by using these ANNs. Baccigalupi \etal (2000)~\cite{Baccigalupi_2000} utilize ANN to separate different types of foregrounds (e.g., thermal dust emissions, galactic synchrotron, and radiation emitted by galaxy clusters) from CMB signal. Petroff \etal (2020)~\cite{Petroff_2020} develop a Bayesian spherical convolutional neural network (CNN) to recover the CMB anisotropies from the foreground contaminations. Using a CNN, Shallue \& Einstein (2023)~\cite{Shallue_2023} reconstruct the initial conditions of the Universe from the late-time density fields which evolve non-linearly. ML can be employed to reconstruct full sky CMB temperature anisotropies from the partial sky maps~\cite{Chanda_2021,Pal_2023a}. Khan \& Saha (2023)~\cite{Khan_2023} develop an ANN to estimate the dipole modulation from the foreground cleaned CMB temperature anisotropies. In our previous article~\cite{Pal_2023b}, we use a CNN to recover full sky $E$- and $B$-modes polarizations from the partial sky maps avoiding the so-called $E$-to-$B$ leakage. Wang \etal (2020b)~\cite{Wang_2020b} use ANN to estimate cosmological parameters from the temperature power spectrum of CMB. Bengaly \etal (2023)~\cite{Bengaly_2023} employ an ML algorithm to constrain the Hubble constant ($H_0$) by using DA Hubble measurements. Moreover, the recent literature~\cite{Liu_2019,Arjona_2020,Mukherjee_2022,Garcia_2023,Wang_2021,Liu_2021} also perform the ML techniques to analyze the low redshift data (i.e., Hubble measurements (DA+BAO), Type Ia Supernovae etc.) for constraining the cosmological parameters of present Universe. The literature~\cite{Shukla_2024} explored how memristive neural networks (MNNs) with leakage and mixed delays achieve finite-time synchronization (FTS) utilizing the state feedback and adaptive control strategies to establish adequate conditions based on the Filippov solution and the Lyapunov functional technique. Recent studies have demonstrated the potential of machine learning in cosmology. Escamilla-Rivera \etal (2020)~\cite{Escamilla_2020} introduced RNN+BNN, combining recurrent and Bayesian neural networks to study dark energy models using supernovae data, reducing computational load while accounting for uncertainties. Tilaver \etal (2021)~\cite{Tilaver_2021} applied LSTM and the Fisher Information Matrix to predict the Hubble parameter within the Chaplygin gas framework, and Salti \etal (2021)~\cite{Salti_2021} investigated the progression of the temperature of the CMB radiation by applying Deep Neural Networks for the modified generalized Chaplygin gas model. The recent work~\cite{Salti_2024} utilized deep learning with a tuned LSTM to predict Hubble parameters from cosmic chronometers. In this paper, we develop ParamANN, an ANN for directly modeling the mapping between $H(z)$ and four parameters of $\Lambda$CDM Universe.

We organize our article as the following. In section~\ref{sec:form}, we describe the relation between the Hubble parameter and redshift for $\Lambda$CDM Universe. In section~\ref{sec:method}, we present the observed $H(z)$ data, discuss the signal and noise models used in our analysis, and describe the deep learning of ParamANN respectively. In this section, we also discuss the MCMC method. In section~\ref{sec:results}, we show the predicted parameter values for the test set as well as for the observed $H(z)$ data. We compare the results predicted by ParamANN with the results obtained by MCMC. Finally, in section~\ref{sec:discuss}, we conclude with the beneficial discussions of our analysis.

\section{Formalism}
\label{sec:form}
Einstein's equations, which can be encapsulated into a compact form using tensor notations, are as follows
\begin{eqnarray}
G_{\mu\nu} &=& -\frac{8\pi G}{c^4}T_{\mu\nu}, \label{eqn:ein}
\end{eqnarray}
where Einstein's tensor $G_{\mu\nu}$ is determined by some suitable second order derivative functions of the metric tensor $g_{\mu\nu}$ with respect to the coordinates. In equation~\eref{eqn:ein}, $G$ denotes the universal gravitational constant, $c$ is the velocity of light in vacuum, and $T_{\mu\nu}$ defines the energy-momentum tensor.

In spherical coordinates, the FRW line element can be written as
\begin{eqnarray}
\textrm{d} s^2 &=& c^2\textrm{d} t^2 - a^2(t)\left[\frac{\textrm{d} r^2}{1-kr^2}+r^2\textrm{d} l^2\right], \label{eqn:metric}
\end{eqnarray}
where $r$, $a(t)$ and $k$ define the radial comoving coordinate, cosmological scale factor and curvature constant of the Universe respectively. The positive, negative, and zero values of $k$ indicate spatially closed, open, and flat Universes respectively. Moreover, $\rm{d} l^2$ is defined as
\begin{eqnarray}
\textrm{d} l^2 &=& \textrm{d} r^2 +\sin^2\theta\textrm{d} \phi^2, \label{eqn:solid_angle}
\end{eqnarray}
where $\theta$ and $\phi$ are the angular comoving coordinates of the spherical Universe.

Using equations~\eref{eqn:ein} and~\eref{eqn:metric}, we obtain two well-known Friedmann equations which are expressed as
\begin{align}
\frac{1}{a^2(t)}\left[\left(\frac{\textrm{d}a}{\textrm{d}t}\right)^2+kc^2\right] &= \frac{8\pi G}{3}\rho(t), \label{eqn:fried1}\\
\frac{2}{a(t)}\frac{\textrm{d}^2a}{\textrm{d}t^2}+\frac{1}{a^2(t)}\left[\left(\frac{\textrm{d}a}{\textrm{d}t}\right)^2+kc^2\right] &= -\frac{8\pi G}{c^2}P(t),\label{eqn:fried2}
\end{align}
where $\rho(t)$ and $P(t)$ denote the density and gravitational pressure of the Universe respectively. Additionally, the equation of state can be written as
\begin{eqnarray}
P(t) &=& \omega c^2\rho(t), \label{eqn:eos}
\end{eqnarray}
where $\omega$ denotes the equation of state parameter. Using equations~\eref{eqn:fried1},~\eref{eqn:fried2} and~\eref{eqn:eos}, we obtain the energy-momentum conservation law which is given by
\begin{eqnarray}
\frac{\partial\rho}{\partial t} +3\left(1+\omega\right)H(t)\rho(t) &=& 0, \label{eqn:conserv_law}
\end{eqnarray}
where $H(t)$ represents the Hubble parameter which is defined by
\begin{eqnarray}
H(t) &=& \frac{1}{a(t)}\frac{\textrm{d}a}{\textrm{d}t}. \label{eqn:Ht}
\end{eqnarray}
The values of $\omega$ are represented as the following
\begin{eqnarray}
\omega &=& \begin{cases}
    \hspace{13pt}0 & \qquad\text{Matter density},\\
    \hspace{13pt}\frac{1}{3} & \qquad\text{Radiation density},\\
    \ -1 & \qquad\text{Vacuum density}.
  \end{cases} \label{eqn:eos_values}
\end{eqnarray}
Using equation~\eref{eqn:conserv_law},~\eref{eqn:Ht} and~\eref{eqn:eos_values}, density components of the Universe are expressed by
\begin{eqnarray}
\rho_{m} &=& \rho_{0m}\left(1+z\right)^3, \label{eqn:mat_den}\\
\rho_{r} &=& \rho_{0r}\left(1+z\right)^4, \label{eqn:rad_den}\\
\rho_{\Lambda} &=& \rho_{0\Lambda}, \label{eqn:vac_den}
\end{eqnarray}
where $z$ denotes the redshift and `0' stands for the present Universe (i.e., $z = 0$). In equations~\eref{eqn:mat_den},~\eref{eqn:rad_den} and~\eref{eqn:vac_den}, subscript notations $m,r$ and $\Lambda$ define the matter, radiation and vacuum densities respectively. Moreover, redshift is related to the scale factor as follows
\begin{eqnarray}
(1 + z) &=& \frac{a_0}{a},
\end{eqnarray}
where $a_0$ denotes today's scale factor of the Universe.

Neglecting radiation density for late time Universe, equation~\eref{eqn:fried1} can be written as
\begin{eqnarray}
\Omega_{m}+\Omega_{k}+\Omega_{\Lambda} &=& 1. \label{eqn:tot_den_par}
\end{eqnarray}
In equation~\eref{eqn:tot_den_par}, $\Omega_m$ and $\Omega_{\Lambda}$ denote the matter and vacuum density parameters respectively, which are defined by
\begin{eqnarray}
\Omega_m &=& \frac{\rho_m}{\rho_c},\label{eqn:mat_den_par}\\
\Omega_{\Lambda} &=& \frac{\rho_{\Lambda}}{\rho_c},\label{eqn:rad_den_par}
\end{eqnarray}
where $\rho_c$ is called the critical density which is expressed as
\begin{eqnarray}
\rho_c &=& \frac{3H^2}{8\pi G}.\label{eqn:crit_den}
\end{eqnarray}
Moreover, the curvature density parameter ($\Omega_{k}$) is defined by 
\begin{eqnarray}
\Omega_{k} &=& -\frac{kc^2}{a^2 H^2}.
\end{eqnarray}

Using equations~\eref{eqn:mat_den} and~\eref{eqn:rad_den} in equation~\eref{eqn:tot_den_par}, the Hubble parameter for $\Lambda$CDM Universe is expressed by
\begin{eqnarray}
H^2(z) &=& H^2_0\left[\Omega_{0m}\left(1+z\right)^3+\Omega_{0k}\left(1+z\right)^2+\Omega_{0\Lambda}\right],\label{eqn:Hz}
\end{eqnarray}
where $\Omega_{0m}$ and $\Omega_{0\Lambda}$ are the matter and vacuum density parameters in the present Universe respectively. Moreover, $\Omega_{0k}$ is today's curvature density parameter which is specified as $-kc^2 /a^2_0H_0^2$. Zero value of $\Omega_{0k}$ specifies a spatially flat Universe. Positive and negative values of $\Omega_{0k}$ define the open and closed Universes respectively. At present Universe, i.e. $z = 0$, equation~\eref{eqn:Hz} provides
\begin{eqnarray}
\Omega_{0m}+\Omega_{0k}+\Omega_{0\Lambda} &=& 1. \label{eqn:today_den_par}
\end{eqnarray}
Using equation~\eref{eqn:today_den_par}, to apply the condition of the present Universe on the curvature density parameter, equation~\eref{eqn:Hz} can be written as
\begin{eqnarray}
H^2(z) &=& H^2_0\Bigl[\Omega_{0m}\left(1+z\right)^3+\bigl(1-\Omega_{0m}-\Omega_{0\Lambda}\bigr)\left(1+z\right)^2+\Omega_{0\Lambda}\Bigr]. \label{eqn:Hz_mod}
\end{eqnarray}
Equation~\eref{eqn:Hz_mod} contains three independent parameters $H_0$, $\Omega_{0m}$, and $\Omega_{0\Lambda}$. However, the curvature density parameter $\Omega_{0k}$ depends on $\Omega_{0m}$ and $\Omega_{0\Lambda}$ by equation~\eref{eqn:today_den_par}. Equation~\eref{eqn:Hz_mod} represents the theoretical model of Hubble parameters at different redshifts for a given set of values of $H_0$, $\Omega_{0m}$, and $\Omega_{0\Lambda}$.

\section{Methodology}
\label{sec:method}
In section~\ref{sec:hubble_data}, we discuss the $H(z)$ measurements obtained by following the DA technique (available in the range $0.07 \leq z \leq 1.965$). Then, in section~\ref{sec:signal} we describe the procedure to simulate the mock values of $H(z)$ and
\begingroup
\setlength{\tabcolsep}{20pt} 
\begin{table}[h!]
\centering
\caption{Table shows the $H(z)$ and corresponding uncertainties, i.e. $\sigma_{H(z)}$, in $\rm{kmMpc^{-1}s^{-1}}$ at different redshifts, measured by DA technique. We consider the covariances for the Hubble data which are shown with bold fonts in this table. We refer the section~\ref{sec:noise} for the detailed discussions about covariances between $H(z)$ data.}\label{table:Hz}
\begin{tabular}{|lllc|}
\hline
$z$ & $H(z)$ & $\sigma_{H(z)}$ & Reference \\
\hline
$0.07$ & $69$ & $19.6$ &~\cite{Zhang_2014} \\
$0.09$ & $69$ & $12$ &~\cite{Jimenez_2003} \\
$0.12$ & $68.6$ & $26.2$ &~\cite{Zhang_2014} \\
$0.17$ & $83$ & $8$ &~\cite{Simon_2005} \\
$\bf{0.1791}$ & $\bf{75}$ & $\bf{4}$ &~\cite{Moresco_2012} \\
$\bf{0.1993}$ & $\bf{75}$ & $\bf{5}$ &~\cite{Moresco_2012} \\
$0.2$ & $72.9$ & $29.6$ &~\cite{Zhang_2014} \\
$0.27$ & $77$ & $14$ &~\cite{Simon_2005} \\
$0.28$ & $88.8$ & $36.64$ &~\cite{Zhang_2014} \\
$\bf{0.3519}$ & $\bf{83}$ & $\bf{14}$ &~\cite{Moresco_2012} \\
$\bf{0.3802}$ & $\bf{83}$ & $\bf{13.5}$ &~\cite{Moresco_2016} \\
$0.4$ & $95$ & $17$ &~\cite{Simon_2005} \\
$\bf{0.4004}$ & $\bf{77}$ & $\bf{10.2}$ &~\cite{Moresco_2016} \\
$\bf{0.4247}$ & $\bf{87.1}$ & $\bf{11.2}$ &~\cite{Moresco_2016} \\
$\bf{0.4497}$ & $\bf{92.8}$ & $\bf{12.9}$ &~\cite{Moresco_2016} \\
$0.47$ & $89$ & $34$ &~\cite{Ratsimbazafy_2017} \\
$\bf{0.4783}$ & $\bf{80.9}$ & $\bf{9}$ &~\cite{Moresco_2016} \\
$0.48$ & $97$ & $62$ &~\cite{Stern_2010} \\
$\bf{0.5929}$ & $\bf{104}$ & $\bf{13}$ &~\cite{Moresco_2012} \\
$\bf{0.6797}$ & $\bf{92}$ & $\bf{8}$ &~\cite{Moresco_2012} \\
$\bf{0.7812}$ & $\bf{105}$ & $\bf{12}$ &~\cite{Moresco_2012} \\
$\bf{0.8754}$ & $\bf{125}$ & $\bf{17}$ &~\cite{Moresco_2012} \\
$0.88$ & $90$ & $40$ &~\cite{Stern_2010} \\
$0.9$ & $117$ & $23$ &~\cite{Simon_2005} \\
$\bf{1.037}$ & $\bf{154}$ & $\bf{20}$ &~\cite{Moresco_2012} \\
$1.3$ & $168$ & $17$ &~\cite{Simon_2005} \\
$\bf{1.363}$ & $\bf{160}$ & $\bf{33.6}$ &~\cite{Moresco_2015} \\
$1.43$ & $177$ & $18$ &~\cite{Simon_2005} \\
$1.53$ & $140$ & $14$ &~\cite{Simon_2005} \\
$1.75$ & $202$ & $40$ &~\cite{Simon_2005} \\
$\bf{1.965}$ & $\bf{186.5}$ & $\bf{50.4}$ &~\cite{Moresco_2015} \\
\hline
\end{tabular}
\end{table}
\endgroup
in section~\ref{sec:noise} we discuss the noise addition to these mock $H(z)$ data. Thereafter, in section~\ref{sec:deep_learn} we provide the descriptions of the deep learning of the neural network used in our analysis. In section~\ref{sec:mcmc}, we describe the MCMC approach which is performed to compare our AI technique.

\subsection{Hubble measurements}
\label{sec:hubble_data}
In our analysis, we create an ANN method to estimate Hubble constant ($H_0$) and density parameters ($\Omega_{0m}$, $\Omega_{0k}$, and $\Omega_{0\Lambda}$) focussing on $31$ available $H(z)$ measured by DA technique~\cite{Jimenez_2002}. These $H(z)$ values are observed in the redshift range $0.07 \leq z \leq 1.965$ without assuming any cosmological model. In table~\ref{table:Hz}, we show these observed $H(z)$ values and corresponding uncertainties in $\rm{kmMpc^{-1}s^{-1}}$ unit. We consider the redshift points of these observed $H(z)$ data to generate the mock values of $H(z)$ for our analysis.

\subsection{Signal model}
\label{sec:signal}
We use \texttt{random.uniform}\footnote{\url{https://numpy.org/doc/stable/reference/random/generated/numpy.random.uniform .html}} function of the Python library \texttt{numpy}\footnote{\url{https://numpy.org/}} to simulate the values of Hubble constant ($H_0$), matter density ($\Omega_{0m}$), and vacuum density ($\Omega_{0\Lambda}$) of the Universe in the suitable ranges. To keep our analysis free from any prejudice that may stem from the restricted choice of priors, we use wide prior ranges for each independent parameter. We consider the uniform range $\left[50, 90\right]$ $\rm{kmMpc^{-1}s^{-1}}$ for $H_0$. Similarly, we utilize the uniform ranges $\left[0.1, 0.7\right]$ and $\left[0.3, 0.9\right]$ for $\Omega_{0m}$ and $\Omega_{0\Lambda}$ respectively. We generate $1.2 \times 10^5$ random values for each of these three parameters in their corresponding uniform ranges. Then, we use equation~\eref{eqn:today_den_par} to obtain the values of $\Omega_{0k}$ by using the simulated values of $\Omega_{0m}$ and $\Omega_{0\Lambda}$. After simulating the values of $H_0$, $\Omega_{0m}$ and $\Omega_{0\Lambda}$, we use the equation~\eref{eqn:Hz_mod} to generate the values of $H(z)$ at $31$ observed redshift points (shown in table~\ref{table:Hz}) for a given set of the simulated values of these cosmological parameters. Finally, we obtain $1.2 \times 10^5$ number of realizations of $H(z)$, where each realization contains $31$ $H(z)$ values corresponding to the observed redshift range $0.07 \leq z \leq 1.965$.

\subsection{Noise model}
\label{sec:noise}
\begin{figure}
\centering
\includegraphics[scale=0.5]{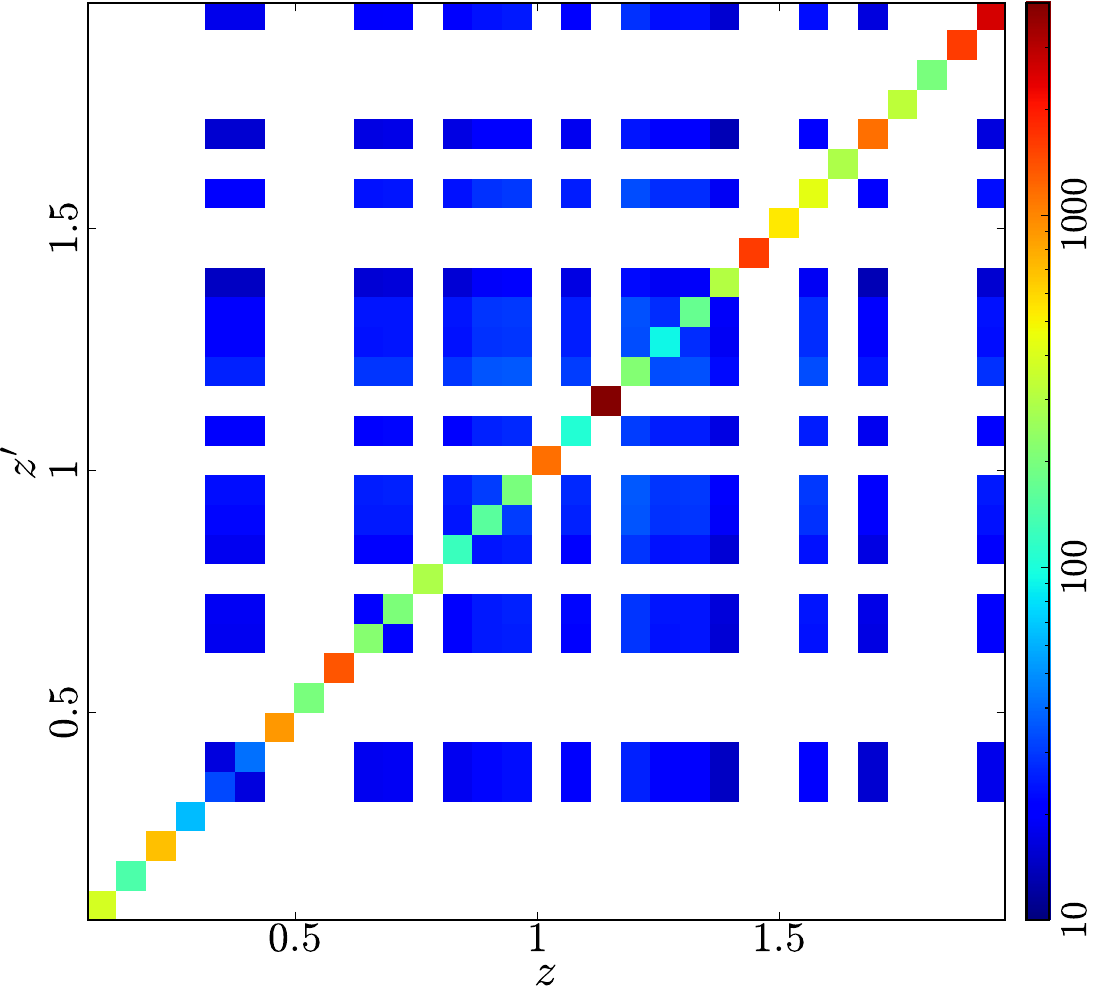}
\caption{Figure shows the full covariance matrix of the observed $H(z)$ in the redshift interval $0.07 \leq z \leq 1.965$. Colorbar represents the non-zero covariances between $H(z)$ measurements in $\log$ scale. White pixels in the covariance plot represent redshift pairs for which covariance informations are not available.}
\label{fig:cov_mat}
\end{figure}
We incorporate correlated Gaussian noises in the simulated $H(z)$ data to account the inherent noises of the Hubble measurements. We generate a full covariance matrix corresponding to the Hubble measurements following the procedure of Moresco \etal (2020)~\cite{Moresco_2020}. In figure~\ref{fig:cov_mat}, we show this full covariance matrix which is used to generate the correlated Gaussian noises. We utilize the openly available code\footnote{\url{https://gitlab.com/mmoresco/CCcovariance}}~\cite{Moresco_2020} to estimate the covariances between Hubble parameters measured by Moresco \etal (2012, 2016)~\cite{Moresco_2012,Moresco_2016} and Moresco (2015)~\cite{Moresco_2015}. Moreover, following the suggestion made by Moresco \etal (2020)~\cite{Moresco_2020}, we consider the bias calculations for the systematic contributions due to the initial mass function (IMF) and stellar population synthesis (SPS) model (odd one out) for the estimation of the covariances corresponding to these Hubble measurements~\cite{Moresco_2012,Moresco_2016,Moresco_2015}. We refer to the literature~\cite{Moresco_2012,Moresco_2016,Moresco_2015,Moresco_2020} for details about the systematic contributions in the Hubble parameters measured by the DA technique. For the rest~\cite{Zhang_2014,Jimenez_2003,Simon_2005,Ratsimbazafy_2017,Stern_2010} of observed Hubble data, we utilize only corresponding variances. In this figure~\ref{fig:cov_mat}, the diagonal values of the full covariance matrix represent the variances of the Hubble parameters measured in the redshift interval $0.07 \leq z \leq 1.965$. Moreover, the off-diagonal region of this covariance matrix represents the covariances between these observed Hubble parameters. The values of variances and covariances are represented by the color bar in the figure~\ref{fig:cov_mat}. We note that the white pixels of this covariance matrix denote the redshift pairs for which no covariance information is available. We consider the noise distributions around $0$ mean, which are bounded by the full covariance matrix (shown in figure~\ref{fig:cov_mat}) corresponding to $31$ Hubble measurements. Using \texttt{random.multivariate\_normal}\footnote{\url{https://numpy.org/doc/stable/reference/random/generated/numpy.random.multivariate\_normal.html}} function of \texttt{numpy} library, we generate $1.2 \times 10^5$ realizations of correlated Gaussian noises by using different seed values. Each of these noise realizations contains $31$ values corresponding to the observed redshift points (shown in table~\ref{table:Hz}). We add these correlated noises to the simulated $H(z)$ data. These noise-included realizations of $H(z)$ are used as the input in ParamANN and the corresponding parameter values are provided as the targets in the output layer of ParamANN.

\subsection{Deep learning of ParamANN}
\label{sec:deep_learn}
We use open-source ML platform TensorFlow\footnote{\url{https://www.tensorflow.org/}}~\cite{Abadi_2015}, using Python programming language, to create the architecture of ParamANN as well as for deep learning of this neural network. In section~\ref{sec:ParamANN}, we describe the model of ParamANN. In section~\ref{sec:preprocess}, we discuss the preprocessing of data and in section~\ref{sec:loss}, we present the loss function used in ParamANN. In section~\ref{sec:train_pred}, we describe the procedure of training and prediction of ParamANN.

\subsubsection{ParamANN}
\label{sec:ParamANN}
\begin{figure}[h!]
\centering
\includegraphics[scale=0.7]{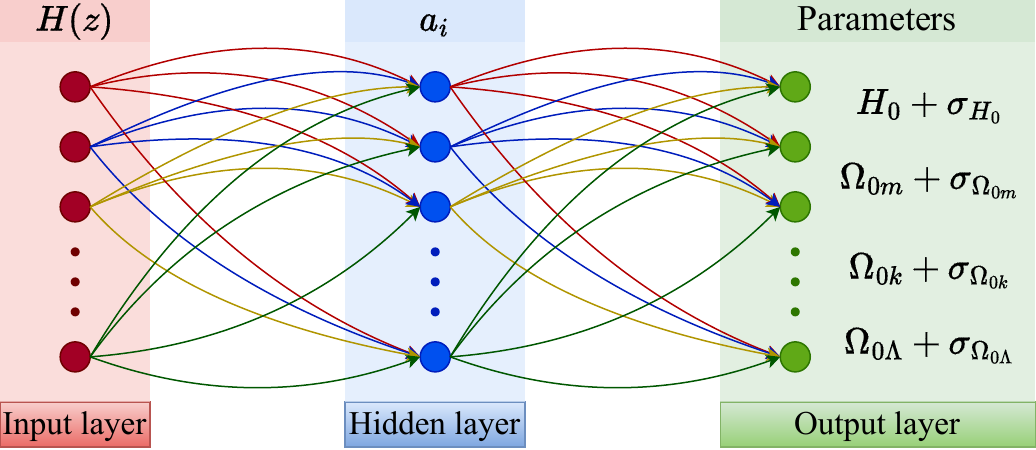}
\caption{Figure shows the architecture of ParamANN, which contains one hidden layer in between input and output layers. Input layer contains $31$ neurons representing $31$ noise-included $H(z)$ values in the range $0.07 \leq z \leq 1.965$, and hidden layer contains $20$ neurons with ReLU activation (i.e., $\left[a_i : i \in \{0, 1, ..., 19\}\right]$). Output layer contains 8 neurons, in which first four neurons represent the predicted parameters (i.e., $H_0$, $\Omega_{0m}$, $\Omega_{0k}$ and $\Omega_{0\Lambda}$) and last four neurons predict the uncertainties corresponding to these parameters.}
\label{fig:ParamANN}
\end{figure}

We construct ParamANN with one hidden layer for the direct mapping between noise-included H(z) and four fundamental parameters, i.e. $H_0$, $\Omega_{0m}$, $\Omega_{0k}$ and $\Omega_{0\Lambda}$, of $\Lambda$CDM Universe. In figure~\ref{fig:ParamANN}, we present the architecture of ParamANN. In ParamANN, the input layer contains thirty one neurons, the hidden layer consists of twenty neurons and the output layer comprises eight neurons. Neurons of each layer are densely connected (by weights and biases) to each neuron of the previous layer. We note that the first half of the output layer shows the predictions and the second half of the same provides the uncertainties corresponding to these predictions. To train the ParamANN, we use the noise-included $H(z)$ as input and the corresponding parameters ($\hat{H}_0$, $\hat{\Omega}_{0m}$, $\hat{\Omega}_{0k}$ and $\hat{\Omega}_{0\Lambda}$) are used as targets, where the `hat' notation is used to define the target parameters. We use ReLU~\cite{Agarap_2019} activation function in the hidden layer to learn the non-linearity of the direct mapping between input and targets. For the optimization process using mini-batch algorithm~\cite{Ruder_2016,Sun_2020}, we utilize \textit{adaptive moment estimation} (ADAM)~\cite{Kingma_2014} optimizer with learning rate $5 \times 10^{-4}$ to update weights and biases in the backward propagation~\cite{Hecht_1992} of the training process of ParamANN.

\subsubsection{Preprocessing of data}
\label{sec:preprocess}
Preprocessing is widely known procedure to normalize the input data (generally distributing the values in the lower range) for the better performance of supervised learning of ANN. Moreover, normalized input data can speed up the training process of neural network. The familiar techniques for the preprocessing are \textit{min-max normalization, z-score normalization} (i.e., \textit{standardization}) etc.~\cite{Kotsiantis_2007}. We use the \textit{standardization} method to normalize the noise-included $H(z)$ data which are used as input in ParamANN. Four fundamental parameters, i.e. $\hat{H}_0$, $\hat{\Omega}_{0m}$, $\hat{\Omega}_{0k}$ and $\hat{\Omega}_{0\Lambda}$, corresponding to these input are used as targets without any scaling. After random shuffling of the entire data, we split these data into three sets, i.e. training, validation and test sets. We use $10^5$ data for training, $1.5 \times 10^4$ data for validation and $5 \times 10^3$ data for testing the predictions of ParamANN. To perform the \textit{standardization} in input, at first, we obtain the mean and standard deviation of noise-included $H(z)$ data of the training set. Then, each sample of these three sets is subtracted by this mean and divided by this standard deviation. We use the mean and standard deviation of the training set even in the validation and test sets, which helps to pass the informations about the training of ParamANN into these data sets effectively. After normalizing the noise-included $H(z)$ of these three sets, by using the standardization method, we use these standardized $H(z)$ data as normalized input in ParamANN.

\subsubsection{Loss function}
\label{sec:loss}
In our analysis, we use \textit{heteroscedastic} (HS) loss function~\cite{Kendall_2017} which is given by
\begin{eqnarray}
L^{HS} &=& \frac{1}{2n}\sum_{q=1}^{n}\left[\exp(-s_q)\left(y_q-\hat{y}_q\right)^2+s_q\right], \label{eqn:loss}
\end{eqnarray}
where $n$ defines the number of targets which is four in our analysis. In equation~\eref{eqn:loss}, $y_q$ and $\hat{y}_q$ represent the predictions and targets respectively. Moreover, in this equation, $s_q$ is the log variances ($\ln\sigma_q^2$) corresponding to the predictions, where $\sigma_q$ denotes the aleatoric uncertainty of the prediction. Using this special type of loss function, we obtain the uncertainties (i.e., aleatoric uncertainty) corresponding to each prediction. Therefore, the output layer of ParamANN contains eight neurons, in which first four neurons estimate the values of the cosmological parameters and the rest of them evaluate the uncertainties corresponding to these parameters.

This HS loss function resembles the negative log-likelihood function which is commonly utilized in the traditional approaches for cosmological parameter estimation, e.g. MCMC method. In spite of this, there is a fundamental difference between the optimization method used in the neural network approach and the likelihood approaches. The Bayesian likelihood function essentially minimizes the noise variance by employing the appropriate likelihood function (which may be Gaussian) to find the best-fit parameter values given the observed data. The loss function used in neural network can not be interpreted as the likelihood function since it represents the difference between the target and prediction based upon training data. The neural network approach in this context is therefore an example of likelihood-free inference~\cite{Alsing_2018}. We demonstrate that using DA Hubble data both the neural network and traditional MCMC approach produce equivalent results. This emphasizes the complementarity of the two methods.

\subsubsection{Training and prediction}
\label{sec:train_pred}
We train ParamANN by using $10^5$ realizations of standardized $H(z)$ and corresponding target parameters, i.e. $\hat{H}_0$, $\hat{\Omega}_{0m}$, $\hat{\Omega}_{0k}$ and $\hat{\Omega}_{0\Lambda}$. The training process continues in an iterative way by minimizing HS loss function used in the neural network. We use $100$ epochs to decide how many times the optimization process should continue. Moreover, each epoch completes with a fixed number of iterations since we chose the mini-batch size of 128. Depending upon the mini-batch size, each iteration takes a subset from the entire training set to minimize the HS loss function (equation~\eref{eqn:loss}). Therefore, one can estimate the number of iterations in each epoch by taking the ratio between the number of training samples and the mini-batch size. In our analysis, the number of iterations in each epoch is 782. We use \textit{model averaging ensemble} (MAE) method~\cite{Lai_2022} to reduce the epistemic uncertainties\footnote{In the deep learning of ANN, epistemic uncertainty exists due to the lack of knowledge in input data as well as the ignorance about the hyperparameters of ANN model.} in the predicted parameters. For the MAE method, we perform the training of ParamANN 100 times (with the same data and the same tuning of hyperparameters) by varying the initialization of weights using 100 randomly selected seed values. The entire training process of these 100 ensembles takes approximately $80$ minutes to execute in an Intel(R) Core(TM) i7-10700 CPU system (contains two threads in each of eight cores) with 2.9 GHz processor speed. We also use $1.5 \times 10^4$ number of validation data at the time of the training process to check any kind of overfitting or underfitting in the minimization of HS loss function and notice no overfitting or underfitting in the training of ParamANN.

After completion of the entire training process, we predict the values of parameters, i.e. $H_0$, $\Omega_{0m}$, $\Omega_{0k}$ and $\Omega_{0\Lambda}$, with corresponding log variances by using $5 \times 10^3$ number (test set) of standardized $H(z)$ data in 100 ensembles of trained ParamANN. We estimate the final predictions of these parameters by taking the mean of 100 ensembles of the predicted parameter values for each realization of the test set. To calculate the uncertainties corresponding to these final predictions, at first we take the exponential of the log variances to obtain the variances corresponding to 100 ensembles of these parameters for each realization of test set. Then, we estimate the mean of 100 ensembles of variances and take the square root of these averaged variances to obtain the uncertainties corresponding to the final predictions of these parameters for each realization of test set.

\subsection{MCMC}
\label{sec:mcmc}
For the MCMC analysis, we employ the Metropolis-Hastings algorithm~\cite{Hastings_1970} for sampling the posterior distributions of three independent parameters $H_0$, $\Omega_{0m}$ and $\Omega_{0\Lambda}$ by considering the prior ranges of these parameters same as those used in our ParamANN method. The MCMC analysis is performed essentially by random walking in the parameter space, minimizing the likelihood function. In this case, the likelihood function is given by
\begin{eqnarray}
\mathcal{L} &=& \exp{-\frac{1}{2}\sum_{i,j=1}^{31}\Delta_{z_i}\left[C^{-1}\right]_{ij}\Delta_{z_j}}, \label{eqn:like}
\end{eqnarray}
where $i$ and $j$ are the dummy indices, and $\Delta_z$ denotes $\left[H(z)-H_{\rm{obs}}(z)\right]$. We note that $H(z)$ defines the Hubble parameters at $31$ observed redshifts calculated by using equation~\eref{eqn:Hz_mod} and $H_{\rm{obs}}(z)$ denotes the Hubble parameters shown in the second column of table~\ref{table:Hz}. Moreover, in equation~\eref{eqn:like}, $C$ denotes the covariance matrix shown in figure~\ref{fig:cov_mat}. Below, we outline the MCMC procedure~\cite{Eriksen_2006,Verde_2003}\\
\hspace*{10pt}1. We start the MCMC chain with a set of values $\left[\Theta_{1}\right]$ in the parameter space, where $\left[\Theta\right]$ stands for three independent parameters ($H_0$, $\Omega_{0m}$ and $\Omega_{0\Lambda}$) and subscript `$1$' implies the initial parameter values. Then, calculate the likelihood $\mathcal{L}_1=\mathcal{L}(\Theta_1)$ by using the equation~\eref{eqn:like}.\\
\hspace*{10pt}2. We choose a random step to obtain a new set of parameter values $\left[\Theta_2\right]$ and compute the corresponding likelihood $\mathcal{L}_2$. The random step is taken by considering the Gaussian probabilities with standard deviations $\left[\sigma_{s}\right]$ in each direction of the previous parameter values.\\
\hspace*{10pt}3a. For $\mathcal{L}_2/\mathcal{L}_1 \geq 1$, we accept the step, i.e. store $\left[\Theta_2\right]$ as the part of MCMC chain, substitute the new parameter values as $\left[\Theta_1\right]\rightarrow\left[\Theta_2\right]$ and then go to the stage 2.\\
\hspace*{10pt}3b. For $\mathcal{L}_2/\mathcal{L}_1 < 1$, generate a uniform random number $x$ within the range $0 \leq x \leq 1$.
\begin{adjustwidth}{2.25em}{0pt}
(i) For $x \geq \mathcal{L}_2/\mathcal{L}_1$, we reject the step, i.e. store $\left[\Theta_1\right]$ as the part of MCMC chain, and return to the stage 2. Moreover, avoid the storing of a set of parameter values more than once.\\
(ii) For $x < \mathcal{L}_2/\mathcal{L}_1$, we accept the step, i.e. follow the same as the stage 3a.
\end{adjustwidth}
\hspace*{10pt}4. Operate multiple chains following the MCMC procedure with different initialization of parameters for generalizing the posterior distribution of each parameter.

We perform the entire MCMC procedure and obtain $8$ chains by initializing the parameters with all combinations of the lower and upper limits of the parameter ranges. Moreover, we utilize $\left[\sigma_s\right]=\left[5,0.1,0.1\right]$ corresponding to $H_0$, $\Omega_{0m}$ and $\Omega_{0\Lambda}$. After obtaining $8$ chains, we ignore the first $50$ samples of each chain for every parameter to avoid the burn-in phase.

\subsubsection{Thinning and mixing}
\label{sec:thinmix}
\begin{figure}[h!]
\centering
\includegraphics[scale=0.48]{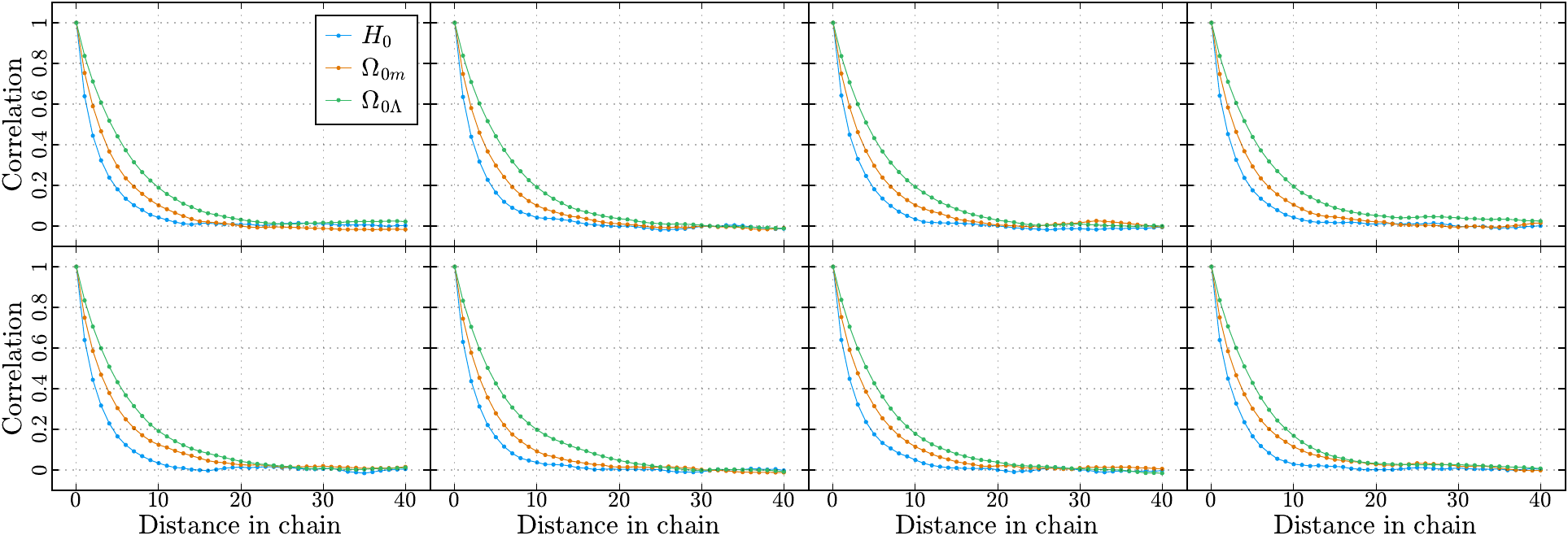}
\caption{Figure contains eight subfigures corresponding to eight different MCMC chains. Each subfigure shows the correlation between samples with respect to the distance in the chain for each of three parameters $H_0$, $\Omega_{0m}$ and $\Omega_{0\Lambda}$. The correlation becomes negligible for the gap of $\sim40$ samples.}
\label{fig:CorrLength}
\end{figure}
\begin{figure}[h!]
\centering
\includegraphics[scale=0.48]{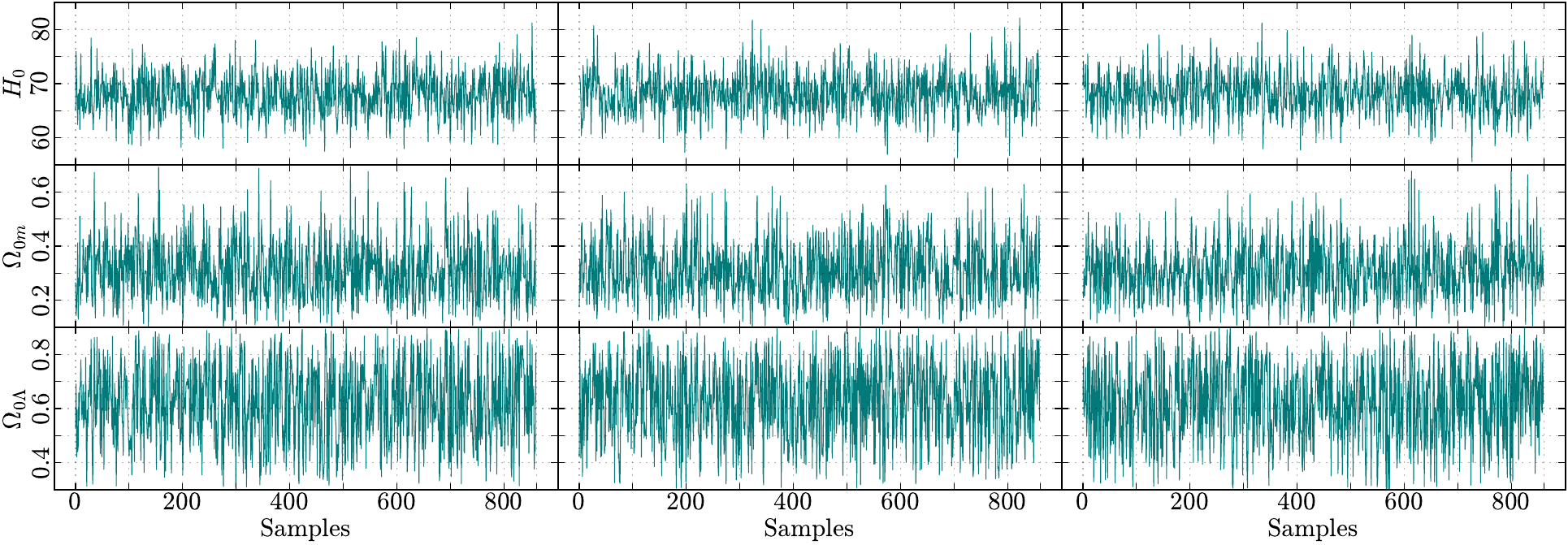}
\caption{Figure shows the trace plots (after performing the thinning process) of the parameters for some of the MCMC chains. Three columns represent three different chains and the rows correspond to three independent parameters as mentioned on the left of each row.}
\label{fig:trace}
\end{figure}
We compute the correlations between samples in every chain for each parameter~\cite{Eriksen_2004}. The correlation function is given by
\begin{eqnarray}
\mathcal{C} (p) &=& \left<\left[\frac{\Theta_i - \left<\Theta\right>}{\sigma_{\Theta}}\right]\left[\frac{\Theta_{i+p} - \left<\Theta\right>}{\sigma_{\Theta}}\right]\right>, \label{eqn:CorrLength}
\end{eqnarray}
where $\left<\cdots\right>$ implies the average operation, $\Theta_i$ denotes the $i$th sample, $\sigma_{\Theta}$ defines the standard deviation of the samples and $p$ represents the distance in chain measured as the number of iterations. In figure~\ref{fig:CorrLength}, we show the correlations between samples for every parameter corresponding to eight chains in eight subfigures. In this figure, the horizontal axis of each subfigure represents the distance in the chain. We notice that the samples of each parameter are sufficiently uncorrelated for the length $p=40$ in each chain. We perform the thinning process using this length and obtain $860$ samples for each parameter in every chain. In figure~\ref{fig:trace}, we present the trace plots of the parameters after thinning in three rows respectively and the columns of this figure represent three different chains. We utilize the entire samples, i.e. $8\times860=6880$, combining all the chains for the best estimate of each parameter. We also obtain $6880$ samples of $\Omega_{0k}$ by using equation~\eref{eqn:today_den_par}.

\subsubsection{Convergence test}
\label{sec:con}
We employ the statistics proposed by Gelman \& Rubin (1992)~\cite{Gelman_1992} to check the convergence of MCMC sampling. The statistical variables are expressed by
\begin{eqnarray}
W &=& \frac{1}{\mathcal{M}(\mathcal{N}-1)}\sum_{i=1}^{\mathcal{M}}\sum_{j=1}^{\mathcal{N}}\left(\Theta_{ij}-\bar{\Theta}_{i}\right)^2, \label{eqn:W}\\
B &=& \frac{\mathcal{N}}{\mathcal{M}-1}\sum_{i=1}^{\mathcal{M}}\left(\bar{\Theta}_{i}-\bar{\Theta}\right)^2,\label{eqn:B}\\
V &=& \left(1-\frac{1}{\mathcal{N}}\right)W + \frac{1}{\mathcal{N}}B,\label{eqn:V}
\end{eqnarray}
where $\mathcal{M}$ denotes the number of chains, $\mathcal{N}$ defines the number of samples in each chain, e.g. $\mathcal{M}=8$ and $\mathcal{N}=860$ for our case. Moreover, $\bar{\Theta}_{i}$ defines the parameter value obtained by averaging over samples for $i$th chain and $\bar{\Theta}$ denotes the parameter value computed by averaging over both samples and chains. For the convergence of MCMC, the ratio $R=V/W$ should be $1$. In our case, we obtain the ratios $0.99983, 0.99923$ and $0.99981$ corresponding to the parameters $H_0$, $\Omega_{0m}$ and $\Omega_{0\Lambda}$ respectively, which ensure the convergence of MCMC sampling of each parameter.

\section{Results and analysis}
\label{sec:results}
In section~\ref{sec:pred_test}, we show the predicted results for the test set. Then, in section~\ref{sec:pred_obs}, we present the predictions of ParamANN and best estimates of MCMC analysis for Hubble measurements. Moreover, we compare the results predicted by ParamANN with the results estimated by both MCMC and Planck collaboration~\cite{Planck_2018}.

\subsection{Predictions for test set}
\label{sec:pred_test}
\begin{figure}[h!]
\centering
\includegraphics[scale=0.47]{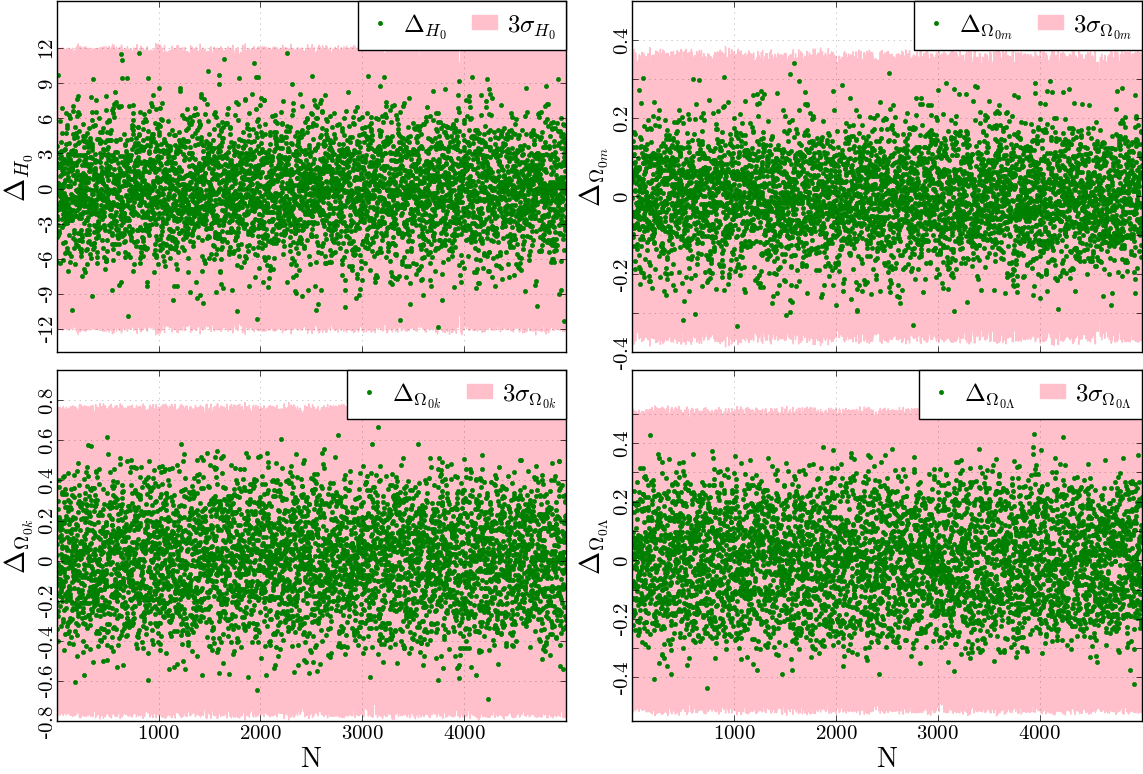}
\caption{Top left subfigure shows the differences between target and predicted Hubble constant. In top right subfigure, we show the differences between target and predicted matter density. Similarly, in bottom left and bottom right subfigures, we present the differences between target and prediction for curvature and vacuum density parameters respectively. Moreover, in each of these subfigures, we show three times uncertainties of the corresponding predictions. Horizontal axis of each subfigure represents the index number of test samples.}
\label{fig:pred_test}
\end{figure}
We predict the Hubble constant ($H_0$) and the density parameters, i.e. $\Omega_{0m}$, $\Omega_{0k}$ and $\Omega_{0\Lambda}$, of $\Lambda$CDM Universe with their corresponding uncertainties, i.e. $\sigma_{H_0}$, $\sigma_{\Omega_{0m}}$, $\sigma_{\Omega_{0k}}$ and $\sigma_{\Omega_{0\Lambda}}$, by using the noise-included $H(z)$ data of test set in trained ParamANN. We compare these predictions with the corresponding targets, i.e. $\hat{H}_0$, $\hat{\Omega}_{0m}$, $\hat{\Omega}_{0k}$ and $\hat{\Omega}_{0\Lambda}$, to show the accuracy of the predictions of ParamANN. We estimate the differences between the targets and predictions of the test set by using the equation which is given by
\begin{eqnarray}
\Delta_{y} &=& \hat{y}-y, \label{eqn:diff}
\end{eqnarray}
where $y$ represents the predicted parameters and $\hat{y}$ defines the corresponding target parameters. In the top left subfigure of figure~\ref{fig:pred_test}, we present the differences between targets and predictions for the Hubble constant. In the same figure, we show similar differences for matter, curvature and vacuum density parameters in top right, bottom left and bottom right subfigures respectively. In each of these subfigures of figure~\ref{fig:pred_test}, we also present three times predicted uncertainties of the corresponding predicted parameters for the test set. We note that the differences corresponding to each of these parameters dominantly lie within their three times uncertainty ranges. Therefore, we conclude that the predictions of test set agree well with the corresponding targets within the corresponding predicted uncertainties.

\subsection{Predictions for Hubble measurements}
\label{sec:pred_obs}
We train the ParamANN by using mock $H(z)$ values (for the range $0.07 \leq z \leq 1.965$) as input. Moreover, these mock $H(z)$ values contain the correlated noises compatible with observed Hubble parameters. Therefore, we can use the observed $H(z)$ values (shown in table~\ref{table:Hz}) directly to the trained ParamANN (after performing the standardization method) to extract the values of Hubble constant, matter, curvature, and vacuum density parameters with their corresponding uncertainties for the present $\Lambda$CDM Universe. Moreover, we perform the MCMC analysis, following the procedure discussed in section~\ref{sec:mcmc}, to obtain the best estimates of the parameters by using Hubble measurements. In section~\ref{sec:hubble_den}, we present the values of Hubble constant and the density parameters with their corresponding uncertainties which are obtained by both trained ParamANN and MCMC analysis for observed $H(z)$ data. In section~\ref{sec:Hz_curve}, we present the Hubble parameter curve using these estimated parameter values comparing with the same obtained by using the results of Planck collaboration~\cite{Planck_2018}.
\begin{table}[h!]
\centering
\caption{Table shows the values of Hubble constant ($H_0$ in $\rm{kmMpc^{-1}s^{-1}}$) and three density parameters ($\Omega_{0m}$, $\Omega_{0k}$ and $\Omega_{0\Lambda}$) obtained by ParamANN, MCMC and Planck collaboration~\protect\cite{Planck_2018} respectively. Last two columns of this table represent the significance of deviation of each of ParamANN and MCMC results from the Planck's results~\protect\cite{Planck_2018}.}\label{table:param_ann}
\begin{tabular}{|l|ccc|cc|}
\hline
\multirow{2}{*}{Parameter} & \multicolumn{3}{c|}{Value} & \multicolumn{2}{c|}{Significance} \\ \cline{2-6} & ParamANN & MCMC & Planck & ParamANN & MCMC \\
\hline
$H_0$ & $68.14 \pm 3.96$ & $68.10^{+4.10}_{-4.16}$ & $67.66 \pm 0.42$ & $0.12\sigma$ & $0.11\sigma$\\[5pt]
$\Omega_{0m}$ & $0.3029 \pm 0.1118$ & $0.2926^{+0.1180}_{-0.1053}$ & $0.3111 \pm 0.0056$ & $0.07\sigma$ & $0.16\sigma$\\[5pt]
$\Omega_{0k}$ & $0.0708 \pm 0.2527$ & $0.0626^{+0.2476}_{-0.2356}$ & $0.001 \pm 0.002$ & $0.28\sigma$ & $0.26\sigma$\\[5pt]
$\Omega_{0\Lambda}$ & $0.6258 \pm 0.1689$ & $0.6416^{+0.1508}_{-0.1746}$ & $0.6889 \pm 0.0056$ & $0.37\sigma$ & $0.31\sigma$\\
\hline
\end{tabular}
\end{table}

\subsubsection{Hubble constant and density parameters}
\label{sec:hubble_den}
\begin{figure}[h!]
\centering
\includegraphics[scale=0.55]{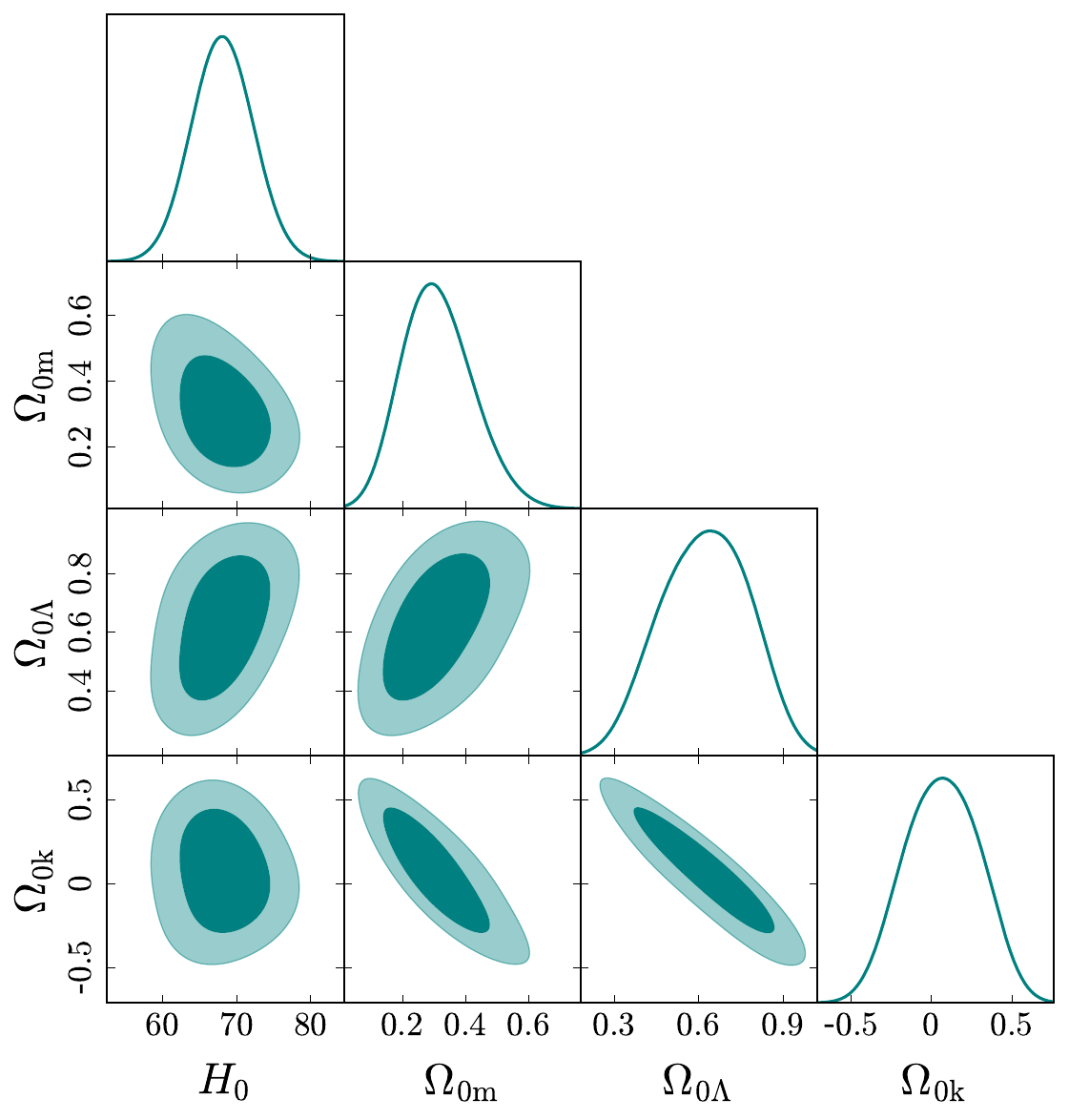}
\caption{Figure shows the interpolated $1\sigma$ and $2\sigma$ regions of 2D joint distribution of the parameters, i.e. $h_0$, $\Omega_{0m}$, $\Omega_{0k}$ and $\Omega_{0\Lambda}$, in the off-diagonal subfigures. The diagonal subfigures represent the interpolated 1D probability density of these parameters.}
\label{fig:mcmc_params}
\end{figure}
In the second and third columns of table~\ref{table:param_ann}, we show the present values of the Hubble parameter ($H_0$) and three density parameters ($\Omega_{0m}$, $\Omega_{0k}$ and $\Omega_{0\Lambda}$) with their corresponding uncertainties estimated by the trained ParamANN and MCMC analysis respectively from the observed Hubble data. We notice that the parameter values estimated by ParamANN and MCMC are equivalent to each other. Even, both ParamANN and MCMC estimate the equivalent uncertainty corresponding to each parameter. In the MCMC analysis, we use the Python library \texttt{GetDist}\footnote{\url{https://getdist.readthedocs.io/en/latest/index.html}} to estimate $68\%$ confidence interval for each parameter. We compare four parameter values estimated by both approaches with the values of these parameters obtained by Planck collaboration~\cite{Planck_2018}. In the fourth column of table~\ref{table:param_ann}, we show the values of these cosmological parameters constrained by Planck collaboration~\cite{Planck_2018} for $\Lambda$CDM Universe. In the cases of ParamANN and MCMC, the estimated uncertainties are comparably larger than the uncertainties of Planck's estimated values of the parameters, since the observed Hubble data are lesser numbers than CMB data as well as these DA $H(z)$ data contain larger uncertainties including various systematic effects. We calculate the significances of deviations of parameter values, for both ParamANN and MCMC, with respect to the Planck's results. We calculate these significances by taking the absolute differences between the estimated parameters and Planck's results as well as dividing these absolute differences by the corresponding uncertainties. We show these significances in the last two columns of table~\ref{table:param_ann} corresponding to ParamANN and MCMC respectively, where $\sigma$ denotes the uncertainties corresponding to the estimated parameters. We notice that the deviation of each parameter from the result of Planck collaboration~\cite{Planck_2018} is significantly low for each of ParamANN and MCMC. In case of ParamANN, the significances are $0.12\sigma$, $0.07\sigma$, $0.28\sigma$ and $0.37\sigma$ corresponding to the parameters $H_0$, $\Omega_{0m}$, $\Omega_{0k}$ and $\Omega_{0\Lambda}$ respectively. Additionally, for the case of MCMC, the significances corresponding to these parameters are $0.11\sigma$, $0.16\sigma$, $0.26\sigma$ and $0.31\sigma$ which are equivalent to the same in case of ParamANN. These low significances (shown in the last two columns of table~\ref{table:param_ann}) corresponding to the estimated parameters indicate the better agreement between ours and Planck's estimates of the parameters $H_0$, $\Omega_{0m}$, $\Omega_{0k}$ and $\Omega_{0\Lambda}$.
\begin{figure}[h!]
\centering
\includegraphics[scale=0.45]{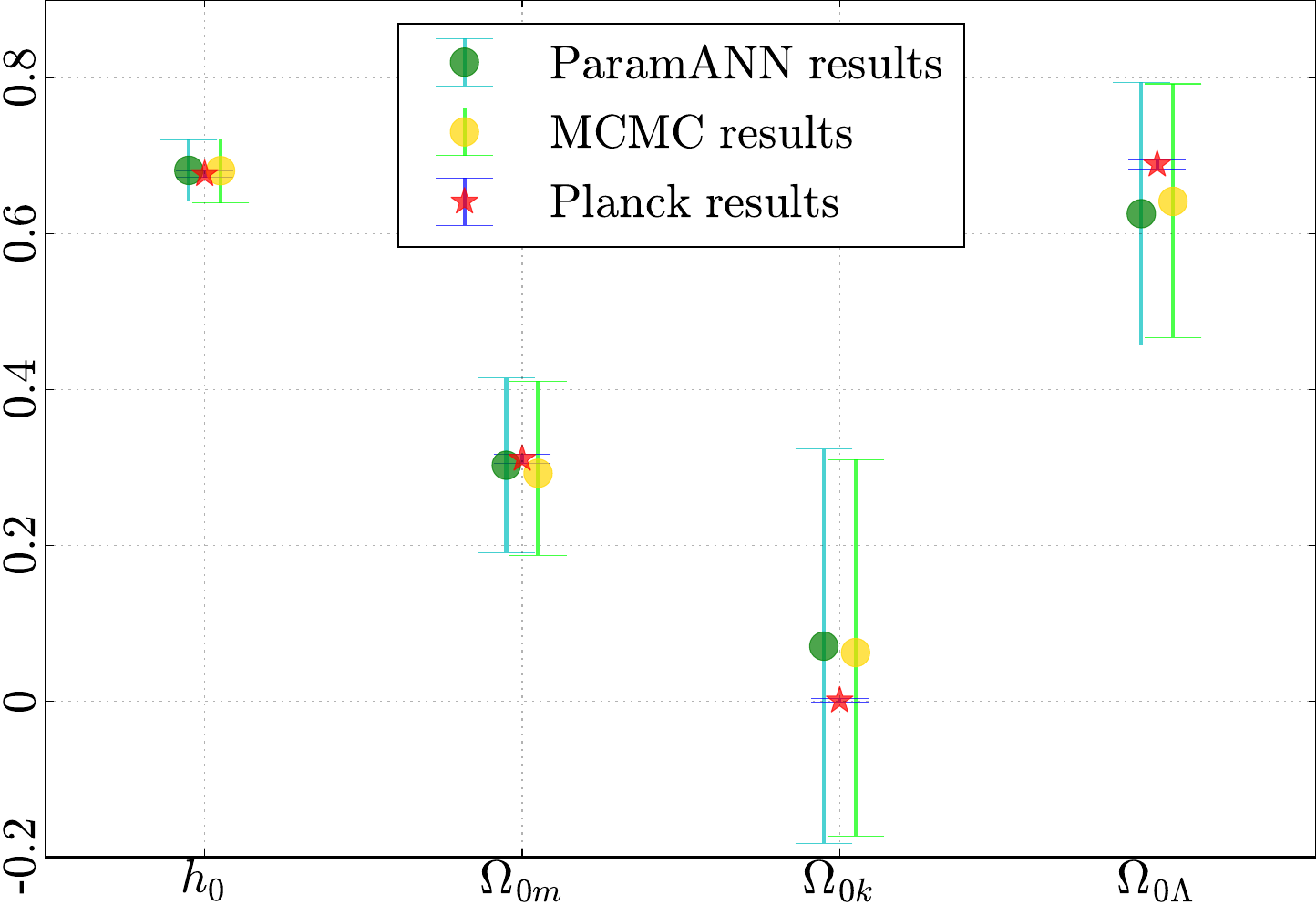}
\caption{Figure represents the values of parameters ($h_0$, $\Omega_{0m}$, $\Omega_{0k}$ and $\Omega_{0\Lambda}$) with corresponding error bars estimated by ParamANN, MCMC and Planck collaboration~\protect\cite{Planck_2018}. The parameter positions corresponding to ParamANN and MCMC are shifted horizontally to differentiate the overlapping points. Hubble constant ($h_0$) is represented in the unit of 100 $\rm{kmMpc^{-1}s^{-1}}$. The horizontal axis indicates the four parameters and the vertical axis represents the values of these parameters.}
\label{fig:params}
\end{figure}

We note that, although the ParamANN is trained for generalized $\Lambda$CDM Universe by considering spatial curvature density, this trained ParamANN predicts spatially flat $\Lambda$CDM Universe by using DA Hubble measurements. Moreover, the parameters predicted by ParamANN are equivalent to the same estimated by MCMC analysis. In figure~\ref{fig:mcmc_params}, we show $1\sigma$ and $2\sigma$ regions of 2D joint distribution of the parameters as well as the marginalized 1D marginalized probability densities of these parameters obtained by using the MCMC samples in \texttt{GetDist}. In~\nameref{sec:appA}, we include the Python code to obtain the figure~\ref{fig:mcmc_params} using \texttt{GetDist}. In figure~\ref{fig:params}, we demonstrate the estimated values of Hubble constant ($h_0$) and three density parameters ($\Omega_{0m}$, $\Omega_{0k}$ and $\Omega_{0\Lambda}$) with corresponding uncertainties provided by trained ParamANN and MCMC analysis. In the same figure, we also present the values of these parameters constrained by Planck collaboration~\cite{Planck_2018}. We horizontally shift the parameter positions corresponding to the ParamANN and MCMC for the visual clarity of the overlapping points. Moreover, in this figure, we present the Hubble constant in 100 $\rm{kmMpc^{-1}s^{-1}}$ unit.

\subsubsection{Hubble parameter curve}
\label{sec:Hz_curve}
\begin{figure}
\centering
\includegraphics[scale=0.45]{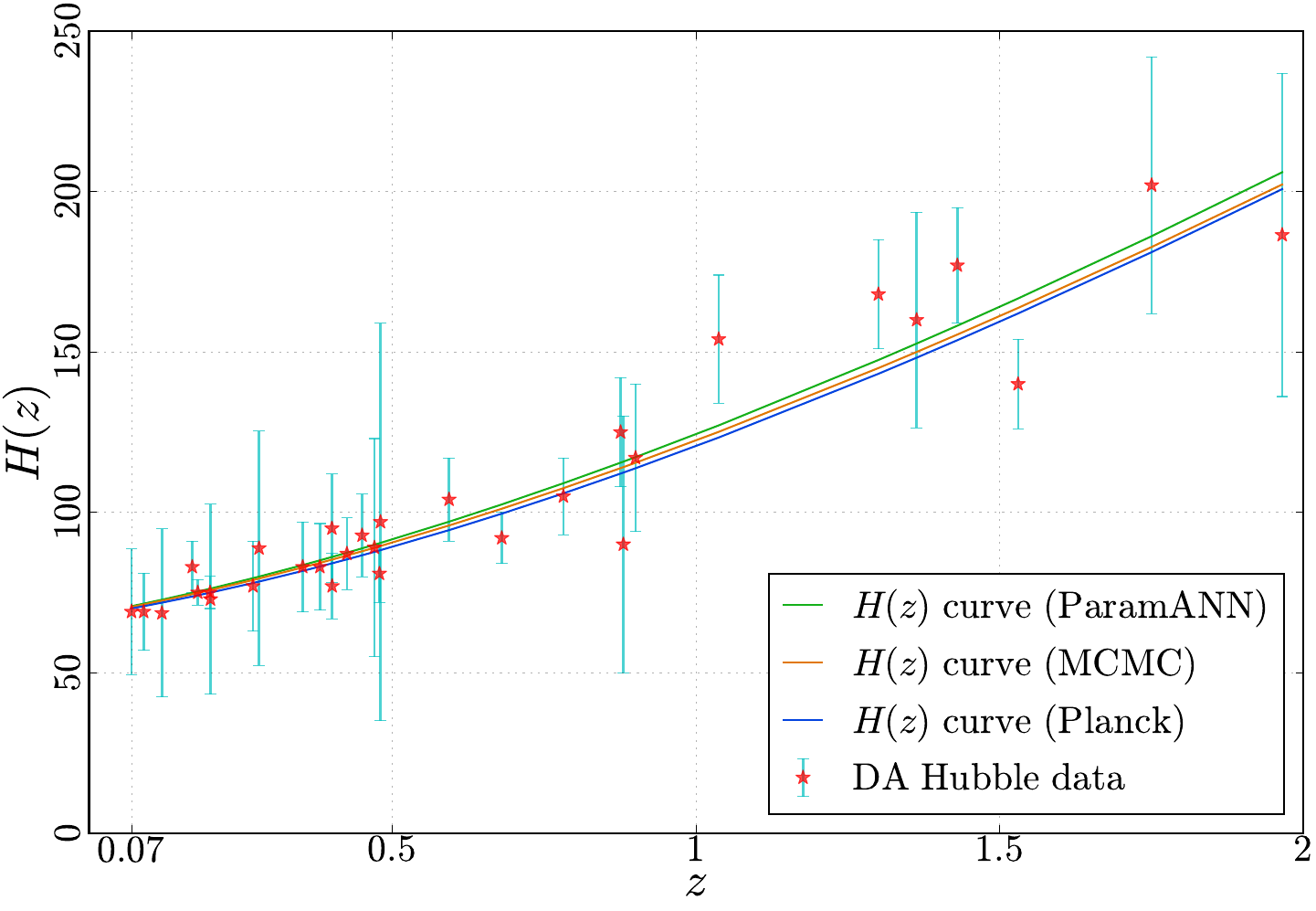}
\caption{Figure shows the Hubble parameter curves computed by using the estimated parameters of ParamANN (green), MCMC (orange) and Planck collaboration (blue)~\protect\cite{Planck_2018} respectively. We also present the Hubble parameters measured by the DA technique. The horizontal and vertical axes of the figure represent the redshift and Hubble parameter (in $\rm{kmMpc^{-1}s^{-1}}$ unit) respectively.}
\label{fig:Hz_curve}
\end{figure}
We obtain the Hubble parameter curves (equation~\eref{eqn:Hz}) using the parameter values (shown in the second and third columns of table~\ref{table:param_ann}) predicted by trained ParamANN as well as estimated by MCMC analysis. We also compute the same by using the parameter values (shown in the fourth column of table~\ref{table:param_ann}) constrained by Planck collaboration~\cite{Planck_2018}. In figure~\ref{fig:Hz_curve}, we show these three curves along with the data points (in the range $0.07 \leq z \leq 1.965$) measured by the DA technique. We refer to~\nameref {sec:appB} for the Python code of figure~\ref{fig:Hz_curve}. Visually each of the three curves in figure~\ref{fig:Hz_curve} seems to fit the observations well. To test how these curves fit the observed points, we estimate the reduced-$\chi^2$ statistics as defined follow\footnote{Although we have not done a $\chi^2$ fitting by ParamANN, we use the reduced-$\chi^2$ fit to estimate the goodness of the fits.},
\begin{eqnarray}
\textrm{reduced-}\chi^2 &=& \frac{1}{\textrm{dof}}\sum_{i,j=1}^{31}\Delta_{z_i}\left[C^{-1}\right]_{ij}\Delta_{z_j}, \label{eqn:chi}
\end{eqnarray}
where $i$ and $j$ are the dummy indices, $\Delta_z$ denotes $\left[H(z)-H_{\rm{obs}}(z)\right]$, and `dof' indicates the number of degrees of freedom, which is $28$ in this case since we use $31$ Hubble data to estimate three independent parameters. In equation~\eref{eqn:chi}, $H_{\rm{obs}}(z)$ defines $31$ observed Hubble data (shown in table~\ref{table:Hz}), $H(z)$ represents the Hubble parameters estimated by using the estimated parameters (or Planck's results) at observed redshift points, and $C$ denotes the covariance matrix shown in figure~\ref{fig:cov_mat}. The reduced-$\chi^2$ value for our ParamANN analysis is $\sim0.512$. Moreover, we obtain a reduced-$\chi^2$ value $\sim0.515$ for MCMC analysis. A similar value, i.e. $\sim0.528$, was calculated for the Planck curve as well. These reduced-$\chi^2$ values corresponding to the cases of ParamANN, MCMC and Planck are equivalent to each other. This shows that the parameters obtained by both ParamANN and MCMC fit the observations well. Further improvements of the error bars of the predicted parameters can be achieved once $H(z)$ observations of lower errors become available.

\section{Discussions and Conclusions}
\label{sec:discuss}
Recent CMB observations indicate that our Universe follows the flat $\Lambda$CDM model dominated by vacuum density~\cite{Planck_2018}. This vacuum density (i.e., the simplest form of dark energy) perhaps accelerates the expansion of our Universe~\cite{Riess_1998,Perlmutter_1999}. The expansion rate of the present Universe can be measured by estimating the value of Hubble constant ($H_0$). Estimation of $H_0$ from the CMB observations shows a significant tension, i.e. so-called Hubble tension~\cite{Planck_2018,Riess_2018,Riess_2019,Riess_2020,Riess_2021,Riess_2022,Wong_2019,Valentino_2021,Jaeger_2022,Brout_2022}, with the value of today's Hubble parameter constrained by local observational data (e.g., Supernovae type Ia measurements).

In this article, we predict the value of the Hubble constant by employing the ML algorithm using the Hubble parameter locally measured by the DA technique. We also measure the cosmological density parameters, i.e. $\Omega_{0m}$, $\Omega_{0k}$ and $\Omega_{0\Lambda}$, of $\Lambda$CDM model without assuming the spatially flat curvature of our present Universe. $H(z)$ data set is a powerful probe of cosmology since these values are related to cosmological parameters in a relatively simple form (equation~\eref{eqn:Hz}). This has the potential of unambiguous interpretations of the analysis results by identifying the underlying cosmological model. Several earlier literature constrain the cosmological parameters under the assumption of a spatially flat Universe~\cite{Macaulay_2013,Raveri_2016,Shafieloo_2012,Sahni_2014,Zheng_2016,
Linder_2003,Shahalam_2015,Leaf_2017,Geng_2018,Gomez_2018,Bengaly_2023,Liu_2019,Arjona_2020,Mukherjee_2022,Garcia_2023}. However, the work of this article estimates all relevant cosmological parameters without assuming the spatial flatness of the Universe. This work, therefore, provides the generalized investigations of observational data to extract the fundamental informations of our Universe. Very interestingly, our measured $H_0$ value is in sharp agreement with the Planck's estimate~\cite{Planck_2018}. Since these values are in contradiction with the $H_0$ value measured by Supernovae data~\cite{Riess_2022} and other local measurements of $H_0$, further attention may be needed in a future article to investigate the nature and the cause of the origin of this tension.

Additionally, we perform the MCMC analysis to estimate the cosmological parameters, i.e. $H_0$, $\Omega_{0m}$, $\Omega_{0k}$ and $\Omega_{0\Lambda}$, of $\Lambda$CDM Universe to compare the parameters predicted by ParamANN. In a classical fitting-the-paremeter approaches (e.g., using a likelihood function), the analyses are performed by estimating the likelihood function of some distribution of observational data as well as initially providing the prior values of all or some of the parameters. The ML approach used by us can be treated as a complementary approach to these and therefore provides a mechanism to test for the robustness of the scientific results with respect to various methods of analysis. The ML approaches uncover the direct mapping between input and targets to learn the hidden pattern between them. Therefore, ML can be used to learn the complex function to extract the cosmological parameters from observational data. The ML and maximum-likelihood approaches are complementary to each other, yet their results agree well with each other. More technically, in our current work, we show the equivalence of the likelihood-free ML method and the MCMC method which is based upon likelihood analysis.

We create an ANN with one hidden layer containing 20 neurons. We call this ANN by a suitable name of ParamANN. We significantly train the ParamANN using $10^5$ samples of mock $H(z)$ values (which contain correlated noises convenient for observed Hubble data) as input and corresponding parameters as targets which are varied uniformly in their specified ranges, i.e. $\left[50, 90\right]$ $\rm{kmMpc^{-1}s^{-1}}$, $\left[0.1, 0.7\right]$ and $\left[0.3, 0.9\right]$ for $H_0$, $\Omega_{0m}$ and $\Omega_{0\Lambda}$ respectively. We note that the mock values of $\Omega_{0k}$ is calculated by $1-\Omega_{0m}-\Omega_{0\Lambda}$. We use another $1.5 \times 10^4$ samples of mock data to validate the training of ParamANN and $5 \times 10^3$ samples to test the performance of trained ParamANN. We notice that the differences between targets and predictions for each parameter dominantly lie within three times uncertainties (predicted by ParamANN) corresponding to the predictions of test set. These results (shown in figure~\ref{fig:pred_test}) of test set show the excellent agreement between targets and predictions of ParamANN (at least within the predicted uncertainties). Finally, we use the trained ParamANN for the predictions of the Hubble constant and the density parameters from the Hubble parameters measured by the DA technique in the redshift interval $0.07 \leq z \leq 1.965$. We obtain $H_0 = 68.14 \pm 3.96$ $\rm{kmMpc^{-1}s^{-1}}$, $\Omega_{0m} = 0.3029 \pm 0.1118$, $\Omega_{0k} = 0.0708 \pm 0.2527$ and $\Omega_{0\Lambda} = 0.6258 \pm 0.1689$. We note that the predicted density parameters show $0.07\sigma$, $0.28\sigma$ and $0.37\sigma$ deviations from Planck's results for matter, curvature and vacuum respectively. Moreover, our predicted Hubble constant (showing $0.12\sigma$ deviation from Planck's result) agrees well with Planck's estimation of $H_0$. Furthermore, we obtain the Hubble parameter curves (figure~\ref{fig:Hz_curve}) by using the estimated parameter values for each of ParamANN, MCMC and Planck collaboration~\cite{Planck_2018}. We notice that these three curves excellently agree to each other and fit well to the observed Hubble data.

The current article presents the first attempt to measure the four fundamental parameters, i.e. $H_0$, $\Omega_{0m}$, $\Omega_{0k}$ and $\Omega_{0\Lambda}$, of $\Lambda$CDM Universe from the DA Hubble measurements by using ML algorithm. In this current analysis, we consider the spatially non-flat $\Lambda$CDM model to train the ParamANN for the estimates of cosmological density parameters along with the Hubble constant. In a future article, we will employ the ML procedure on the different types of dark energy models (i.e., wCDM, CPL~\cite{Chevalier_2001,Linder_2003}, scalar field etc.) regarding the estimations of fundamental cosmological parameters to compare these dark energy models each other.

\ack We acknowledge the use of open-source software library TensorFlow\footnote{\url{https://www.tensorflow.org/}}, Python libraries \texttt{numpy}\footnote{\url{https://numpy.org/}}, \texttt{GetDist}\footnote{\url{https://getdist.readthedocs.io/en/latest/index.html}} and openly available code\footnote{\url{https://gitlab.com/mmoresco/CCcovariance}} given by Moresco. We thank Albin Joseph and Md Ishaque Khan for useful discussions associated with this work.

\section*{Data availability statement}
The data can not be made publicly available upon publication because no suitable repository exists for hosting data in this field of study. The authors will share the data of this work after receiving a reasonable request.

\appendix

\section*{Appendix A}
\label{sec:appA}

We use the following Python code to obtain the figure~\ref{fig:mcmc_params}.\\
\rule{\columnwidth}{0.4pt}
\texttt{
import getdist\\
from getdist import plots\\
import matplotlib.pyplot as plt\\
\%matplotlib inline\\
plt.style.use(\textquotesingle classic\textquotesingle)\\
samples = getdist.MCSamples(samples=samples\_array,\\
\hspace*{2cm} names=[r\textquotesingle\$H\_0\$\textquotesingle,r\textquotesingle\$\textbackslash Omega\_\{\textbackslash rm\{0m\}\}\$\textquotesingle,\\
\hspace*{3.5cm} r\textquotesingle\$\textbackslash Omega\_\{\textbackslash rm\{0\textbackslash Lambda\}\}\$\textquotesingle,\\
\hspace*{3.5cm} r\textquotesingle\$\textbackslash Omega\_\{\textbackslash rm\{0k\}\}\$\textquotesingle],\\
\hspace*{2cm} settings=\{\textquotesingle smooth\_scale\_1D\textquotesingle:0.75,\textquotesingle smooth\_scale\_2D\textquotesingle:0.75\})\\
g = plots.get\_subplot\_plotter()\\
g.settings.axes\_fontsize=20\\
g.settings.axes\_labelsize=25\\
g.triangle\_plot(samples,filled=True,line\_args=\{\textquotesingle lw\textquotesingle:1.5,\textquotesingle ls\textquotesingle:\textquotesingle -\textquotesingle,\\
\hspace*{3cm} \textquotesingle color\textquotesingle:\textquotesingle teal\textquotesingle\},contour\_colors=[\textquotesingle teal\textquotesingle])\\
ax1 = g.subplots[1, 0]\\
ax2 = g.subplots[2, 0]\\
ax3 = g.subplots[3, 0]\\
ax4 = g.subplots[3, 1]\\
ax5 = g.subplots[3, 2]\\
ax6 = g.subplots[3, 3]\\
g.rotate\_yticklabels(ax=4,rotation=90,labelsize=None)\\
g.rotate\_yticklabels(ax=8,rotation=90,labelsize=None)\\
g.rotate\_yticklabels(ax=12,rotation=90,labelsize=None)\\
ax1.set\_yticks([0.2,0.4,0.6],labels=[r\textquotesingle\$0.2\$\textquotesingle,r\textquotesingle\$0.4\$\textquotesingle,r\textquotesingle\$0.6\$\textquotesingle])\\
ax2.set\_yticks([0.4,0.6,0.8],labels=[r\textquotesingle\$0.4\$\textquotesingle,r\textquotesingle\$0.6\$\textquotesingle,r\textquotesingle\$0.8\$\textquotesingle])\\
ax3.set\_yticks([-0.5,0,0.5],labels=[r\textquotesingle-\$0.5\$\textquotesingle,r\textquotesingle\$0\$\textquotesingle,r\textquotesingle\$0.5\$\textquotesingle])\\
ax3.set\_xticks([60,70,80],labels=[r\textquotesingle\$60\$\textquotesingle,r\textquotesingle\$70\$\textquotesingle,r\textquotesingle\$80\$\textquotesingle])\\
ax4.set\_xticks([0.2,0.4,0.6],labels=[r\textquotesingle\$0.2\$\textquotesingle,r\textquotesingle\$0.4\$\textquotesingle,r\textquotesingle\$0.6\$\textquotesingle])\\
ax5.set\_xticks([0.3,0.6,0.9],labels=[r\textquotesingle\$0.3\$\textquotesingle,r\textquotesingle\$0.6\$\textquotesingle,r\textquotesingle\$0.9\$\textquotesingle])\\
ax6.set\_xticks([-0.5,0,0.5],labels=[r\textquotesingle-\$0.5\$\textquotesingle,r\textquotesingle\$0\$\textquotesingle,r\textquotesingle\$0.5\$\textquotesingle])\\
g.export(\textquotesingle filename\textquotesingle)\\
}
\rule{\columnwidth}{0.4pt}
where \texttt{samples\_array} denotes the array of $6880$ samples of four parameters and \mbox{\texttt{filename}} defines the name of file to save the figure.

\section*{Appendix B}
\label{sec:appB}
The following is the Python code to obtain the figure~\ref{fig:Hz_curve}.\\
\rule{\columnwidth}{0.4pt}
\texttt{
import numpy as np\\
data = np.loadtxt(\textquotesingle hubble.csv\textquotesingle, delimiter=\textquotesingle,\textquotesingle,usecols=(0,1,2))\\
z = data[:,0]\\
Hz = data[:,1]\\
Hz\_std = data[:,2]\\
import matplotlib\\
import matplotlib.pyplot as plt\\
import matplotlib.gridspec as gridspec\\
\%matplotlib inline\\
plt.style.use(\textquotesingle classic\textquotesingle)\\
ann\_Hz = 68.14*np.sqrt((0.3029*((1+z)**3))+(0.0708*((1+z)**2))+0.6258)\\
mcmc\_Hz = 68.10*np.sqrt((0.2926*((1+z)**3))+(0.0626*((1+z)**2))+0.6416)\\
planck\_Hz = 67.66*np.sqrt((0.3111*((1+z)**3))+(0.001*((1+z)**2))+0.6889)\\
fig = plt.figure(figsize=(10,7))\\
plt.errorbar(z,Hz,yerr=Hz\_std,fmt=\textquotesingle r*\textquotesingle,ecolor=\textquotesingle c\textquotesingle,mec=\textquotesingle r\textquotesingle,
markersize=8,\\
\hspace*{2.5cm}elinewidth=1.2,label=r\textquotesingle\$\textbackslash rm\{DA \textbackslash Hubble \textbackslash data\}\$\textquotesingle,alpha=0.7)\\
plt.plot(z,ann\_Hz,\textquotesingle-\textquotesingle,color=\textquotesingle xkcd:green\textquotesingle,linewidth=1,\\
\hspace*{2.5cm}label=r\textquotesingle\$H(z) \textbackslash \textbackslash rm\{curve \textbackslash (ParamANN)\}\$\textquotesingle)\\
plt.plot(z,mcmc\_Hz,\textquotesingle-\textquotesingle,color=\textquotesingle xkcd:pumpkin\textquotesingle,linewidth=1,\\
\hspace*{2.5cm}label=r\textquotesingle\$H(z) \textbackslash \textbackslash rm\{curve \textbackslash (MCMC)\}\$\textquotesingle)\\
plt.plot(z,planck\_Hz,\textquotesingle-\textquotesingle,color=\textquotesingle xkcd:blue\textquotesingle,linewidth=1,\\
\hspace*{2.5cm}label=r\textquotesingle\$H(z) \textbackslash \textbackslash rm\{curve \textbackslash (Planck)\}\$\textquotesingle)\\
plt.grid(alpha=0.3)\\
plt.xticks([0.07,0.5,1,1.5,2],\\
\hspace*{2.5cm}labels=[r\textquotesingle\$0.07\$\textquotesingle,r\textquotesingle\$0.5\$\textquotesingle,r\textquotesingle\$1\$\textquotesingle,r\textquotesingle\$1.5\$\textquotesingle,r\textquotesingle\$2\$\textquotesingle])\\
plt.yticks([0,50,100,150,200,250],\\
\hspace*{2.5cm}labels=[r\textquotesingle\$0\$\textquotesingle,r\textquotesingle\$50\$\textquotesingle,r\textquotesingle\$100\$\textquotesingle,r\textquotesingle\$150\$\textquotesingle,r\textquotesingle\$200\$\textquotesingle,r\textquotesingle\$250\$\textquotesingle])\\
plt.tick\_params(axis=\textquotesingle x\textquotesingle,labelsize=20,pad=0.5)\\
plt.tick\_params(axis=\textquotesingle y\textquotesingle,labelsize=20,pad=0,rotation=90)\\
plt.xlabel(r\textquotesingle\$z\$\textquotesingle,fontsize=25,labelpad=-5)\\
plt.ylabel(r\textquotesingle\$H(z)\$\textquotesingle,fontsize=25,labelpad=0)\\
plt.legend(loc=\textquotesingle lower right\textquotesingle,fontsize=20,numpoints=1,handlelength=1.0)\\
plt.savefig(\textquotesingle filename\textquotesingle,dpi=200,bbox\_inches=\textquotesingle tight\textquotesingle,pad\_inches=0.02)\\
}
\rule{\columnwidth}{0.4pt}
where three columns of \texttt{hubble.csv} data file contain observed redshifts (dimensionless), Hubble parameters and corresponding standard deviation in $\rm{kmMpc^{-1}s^{-1}}$ unit respectively. Moreover, \texttt{filename} represents the name of file to save the figure.

\section*{References}

\end{document}